# The *Biological Anthropocene:* rethinking novelty organisms, interactions, and evolution

Pablo José Francisco Pena Rodrigues[1*]

Catarina Fonseca Lira[1]

[1]Instituto de Pesquisas Jardim Botânico do Rio de Janeiro, Rua Pacheco Leão, 915, CEP 22460-030, Rio de Janeiro, Brazil

*Corresponding author: Rodrigues, P.J.F.P. (pablojfpr@hotmail.com)

**Abstract**

Anthropogenic changes of the biota and human hyper-dominance are modulating the evolution of life on our planet. Humankind has spread worldwide supported by cultural and technological knowledge, and has already modified uncountable biological interactions. While numerous species have been extinguished by human actions, others are directly favored, such as alien species, hybrids, and genetically modified organisms. These biodiversity shifts have generated new interactions among all living organisms in anthropized or anthropogenic ecosystems, with the consequent establishment of novel evolutionary pathways. Thus, humans have created a strong evolutionary bias on Earth, leading to unexpected and irreversible outcomes. Anthropogenic changes and novelty organisms are shifting the evolutionary paths of all organisms towards the *Biological Anthropocene*, a new concept of our imprint on biodiversity and evolution.

**Human hyper-dominance and "future" outcomes**

Human beings have drastically impacted the Earth's surface and promoted striking ecosystem and biodiversity alterations [1,2]. Habitat destruction and pollution, species extinction, biotic homogenization [3], and gene exchange between species [4] are some of the many ways nature is currently changing. Despite apparent biological impoverishment, however, humans could be directly increasing biodiversity [5–7]. In fact, anthropogenic ecosystems, such as cities, may drive evolution and create new organisms [8] – thus establishing new evolutionary pathways.

According to the Modern Synthesis, the main evolutionary forces have been natural selection, genetic drift, gene flow, and mutation [9], although other processes, either natural or induced by humans (such as developmental bias, plasticity, inclusive inheritance, and niche construction), can contribute to the evolution of species, as well as the Extended Evolutionary Synthesis [10].

Anthropogenic changes on Earth and of the evolutionary pathways of biodiversity are occurring in unprecedented ways due to the human hyper-dominance as a 'hyper-keystone' species (sensu [11]). Therefore, in this new Epoch that is modulated by human culture and technology, a drastic reduction of the current biodiversity is expected to occur, followed by the expansion of anthropogenically-favored organisms in all habitats – called the *Biological Anthropocene (BioAnthro).*

The future is not accurately predictable and we cannot anticipate if and how these new evolutionary trends will persist and transform biodiversity through time. Most current evolutionary pathways, however, have already been shifted because of anthropic changes on Earth. We therefore anticipate that alien species, hybrids, and genetically modified organisms (GMOs) will create new evolutionary pressures on all ecosystems, species distributions, and local biodiversity – leading to novel alternate evolutionary pathways.

**The ultimate biological changes and the Anthropocene**

There is little doubt that human activities are causing permanent changes on Earth [12–14], including numerous extinctions [15,16] and the creation of new ecosystems – known as anthropogenic biomes (e.g., Anthromes; [17]). Cultural values and other social structures lead to behavioral choices by individuals and their social groups and induce changes in the environment [18] and also lead to increasing human population pressures on Earth's ecosystems [19].

Some anthropogenic changes have been enormous, such as habitat suppression [20] and the production of ~30 trillion tons (Tt) of technosphere materials and artifacts (see [21]). There are, however, numerous less-noticeable environmental impacts, such as the high production of pesticides [22], fertilizers [23], acid effluents [24], radioactive waste [25], antimicrobial compounds [26], alien species [27,28], GMOs [4], and many others.

Those modifications greatly disturb world ecosystems, and the main consequences we are seeing now are the advent of mass extinctions in the Anthropocene defaunation [29,30] and biotic homogenization [3]. Such processes are observed in many sites where alien and ruderal species become established, especially within cities [31,32]. Briefly, those changes could allow the establishment and persistence of novel organisms through hybridization, genetic drift, and/or selection, where organisms better adapted to anthropogenic conditions would be favored.

Extinctions and habitat changes are occurring in uncontrollable ways with unexpected trajectories [33,34]. This trend may be thought as approaching the "tipping point", where evolutionary patterns are permanently changed by anthropogenic pressures and biological thresholds are definitely crossed [35,36], but since some global characteristics such as functional diversity, novel organisms and atmospheric pollution cannot yet be fully factored in [34,37], the results are unpredictable. The *BioAnthro* concept incorporates multi-scaled, fractal changes in additive patterns, inevitably altering the direction of evolution.

The evolutionary scenarios of the *BioAnthro* demand that organisms adapt more rapidly than they otherwise would in response to ecosystem changes driven by the creation of semi-natural habitats (e.g., [38]), agricultural fields [39], anthropogenic biomes [17], and urban [40] and novel ecosystems [41]. In this sense, organisms capable of fast adaptation, such as alien species, GMOs, economically- and anthropogenically-favored organisms (e.g., crops and livestock animals), or hybrids showing high fitness, will prevail in the *BioAnthro*. These organisms will therefore not only persist, but be favored in new evolutionary pathways because of their resilience and high capacity to adapt to future modified scenarios.

**The concept of the *Biological Anthropocene***

The *BioAnthro* is a concept based on increasing biodiversity and the subsequent creation, spread and transformation of new life forms induced and favored by human activities, which represent the most important human imprint on evolution. The new, or transformed, biological entities are largely influenced by human-driven processes such as hybridization, artificial selection, positive selection for plagues and parasites, environmental transformation, alien species establishment, and the spread and exchange of genes. Thus, the *BioAnthro* is modulated by uncountable new interactions and relationships among already established organisms, newly established novel organisms, humankind itself, and technology –in feedback loops reshaping the evolutionary processes on Earth.

Evolution is a modifying, transforming, and changeable force that depends on the interactions of species [42]. It is therefore natural that evolutionary pathways are constantly changing due to human-driven modifications [43] and the creation of new interactions between organisms – eventually leading to a point of no return (*sensu* [44]). Those changes, either driven, or randomly-caused by humans are immediately imprinted on all living organisms through modified habitats, novel or lost functions, and new interactions [41,45,46].

Many scientists have contradictory views of anthropogenic changes. Some believe they are positive, as many organisms are favored by artificial selection; other scientists have a negative view, with Earth nearing its sixth mass extinction [47].Yet one point is irrefutable – the Anthropocene biota will be different from the current biota in unpredictable and unexpected ways. The Anthropocene is a time of disruptive processes on a planet has already been fundamentally altered by humans [48].

Human hyper-dominance is likely the main driving force of the *BioAnthro*. All existing life forms on Earth may already have shifted to totally unforeseen and alternate evolutionary paths because of indirect or direct anthropogenic pressures. But humans are also one of the main species affected by Anthropocene changes, transforming and simultaneously being transformed. The influence of humans on other organisms (and vice-versa) is not exclusively related to environmental factors but is also driven by technological, social, and cultural evolution [40]. It is very important to recognize that Human-Environment interactions can spread on wide spatial scales [34].

### *BioAnthro*: facts and unpredictability

The *BioAnthro* is largely driven by chance, and by constant human-induced modifications. This new concept of the Human-Environment relationship can be perceived by anyone aware of the anthropogenic changes on Earth, although general global awareness is still incipient. In the life sciences, there are factors that are "known unknowns" as well as "unknown unknowns" concerning the future biota that weave uncertainties into any predictions [49]. However, there are already strong signs that novel organisms and novel interactions are definitely altering evolution.

It is currently hard to identify environmental modifications and species' interactions that are not driven, intentionally or by chance, by human culture and technology [50]. Humans occupy the entire planet, without any special distinctions of habitats, so that all of them have been touched by mankind [51]. Human interventions in the environment also lead to many uncertainties, and have always been associated with risks and consequences. Conservation efforts as we know them today, for example, are unpredictable altering evolutionary processes in nature. Intentional conservation actions favoring a few selected organisms (such as flagship species) change population patterns within the ecosystem [52]. Future outcomes predicted by scientific studies may never come about because of the constant and unpredictable changes on interactions between living organisms and their environment.

From the times of our ancestors (including Neanderthals) to modern *Homo sapiens*, every action, decision, necessity, and even ideas have changed the outcome of evolution. Thus, according to the "Early Anthropogenic hypothesis" (see [46]), the *BioAnthro* may have started thousands of years ago, altering the environment and the evolutionary paths of all species. Humans have potentialized the capacity of a single species to change the environment and evolutionary paths through our hyper-dominance, technological development, and diverse cultural processes [11]. These characteristics lead to an increased unpredictability of the *BioAnthro* due to feedback loops of Human-Environment relationships.

Some examples of the anthropogenic changes causing biodiversity instability and unexpected outcomes are: trophic cascades and predator-prey interactions [53,54], pollinator population declines [55], and spreading parasite vectors [56]. It is likely that population declines and species extinctions have already caused irreversible changes in the dynamics of most ecosystems, and eventually these changes will reach all corners of the planet without distinction. Our current scientific knowledge is much too deficient to understand or predict the types of mutable changes of species' interactions that will occur – including human habits, culture and long-term technological progress.

Besides the changes and environmental shifts caused by human actions, mankind has also added novel organisms created directly or indirectly by processes such as artificial selection (see [57,58]), hybridization (artificial or favored in nature)[59], ploidy changes in animals [60] and plants [61], and new transgenic organisms [4]. Many other changes in species compositions (biotic) [62] and in environment conditions (abiotic) have been influenced by human actions [63–65].

Alien species, hybrids, and GMOs are the main evolutionary novelty organisms currently integrating the new webs of interactions between all living organisms, so that humans have directly and indirectly pushed unexpected evolutionary changes towards the *BioAnthro*. Alien species have a major role in redefining biotic and abiotic conditions in anthropogenic-influenced habitats, especially due to their invasiveness, resilience, high capacity for adaptation, and capacity for rapid evolutionary alteration (see [66]). Many efforts focusing on the eradication of alien species are inefficient because of their high fitness and resilience [67]. Alien species will therefore definitely be present, and even abundant, in many ecosystems, and represent an important factor in the new interactions among living organisms and in the novel evolutionary pathways in the *BioAnthro* (BOX 1).

Within a classical conservation perspective, it is crucial to maintain species integrity to avoid hybridization [68]. The propagation of hybrids in the ecosystem will consequently cause declines in the parental species, which is considered unnatural and a cause for ecosystem management [69]. But there are many examples of human-induced hybrids that are positive and successful [70,71] and that may change the evolutionary outcomes of essentially all living species. It is therefore very likely that hybrids will be favored in future anthropogenic systems, causing even more changes in environmental conditions and interactions (BOX 2).

International debates concerning GMOs show two clearly opposite perspectives, although both have in common anthropocentric points of view. While GMO supporters believe that we need them to serve mankind, others focus on the environmental and uncertainties about health risks [72,73]. GMOs are now so widespread that almost all of our principal food crops are now transgenic [74], with annual productions of billions of transgenic organisms in permeable anthropogenic ecosystems that are likely to allow interactions with the native biota. Not only transgenic GMOs, but other organisms such as cisgenic plants and epicrops, are novelty organisms that will likely influence evolution. There are still many uncertainties surrounding GMOs, and the *BioAnthro* is heavily affected by the widespread presence of genetically-modified organisms coexisting and interacting with other living species (BOX 3).

**Box 1:** *Alien species*

Alien species are those that are non-native to a given ecosystem and (like humans in the Anthropocene) some invasive alien species grow into large populations and potentialize biotic interactions with native organisms [66,67]. It is therefore common sense that alien species represent a threat to local biodiversity once they become widespread and established due to intentional or unintentional human actions [75]. However, in highly anthropized and constantly changing ecosystems, alien species can turn into "survivors". They are often directly-favored and constantly modified by human activities through artificial selection and domestication [76]. Alien species are also intentionally cultivated in areas where they and their wild relatives are not native, such as in the case of crops and livestock animals.

Many studies have shown that alien species not only colonize new habitats but modify those they have invaded (e.g., [66,67,77]), reducing global biodiversity [27,28]. Alien species can generally overspread easily due to human activities [67,78]. Since humans are widespread on Earth, alien organisms have reached and affected most (eventually all) regions of the planet, causing perturbation patterns that might be compared to past mass extinctions [79]. As a consequence, there may be huge biotic homogenizations [3,27], with most anthropized habitats having similar pools of species.

It is also expected that these new interactions and the characteristics of the new invaded habitat will be more suitable to alien organisms. In fact, some newly invaded habitats show large alien populations, mostly in highly anthropized sites. Alien species colonize altered sites even in Protected Areas, although pristine areas seems to be less vulnerable to biotic invasions (e.g., [80]). Alien species can therefore represent not only a disruption of natural processes, but also a new way to resist and even benefit from human impacts. In that sense, it is very plausible to hypothesize that invasive alien species, as compared to native organisms, have greater chances of influencing the new evolutionary pathways of the *BioAnthro*.

**Box 2:** *Hybrids*

    Hybridization is an important evolutionary process for species diversification [59]. The results of that creation of new organisms, and ultimately new taxa, can be positive (with species diversification and the development of domesticated plants and animals), or negative (with the development and evolution of disease vectors, diseases, and pest species) [73,81].

    Nowadays, hybridization is boosted by technology and human habits. Little is known about the effects or the frequencies of horizontal genetic material transfer between related (or even unrelated) species [82], especially in anthropogenic systems such as urban and agricultural areas. Regardless of the complexity of the parental species, most successful and established hybrid organisms have higher fitness than the parental organisms [83]. Even if fitness is not favored, models have shown that hybrids can naturally and rapidly evolve into a new species through reproductive isolation driven by genetic incompatibilities [84]. Additionally, hybrids may evolve differently in anthropogenic-related hybrid zones than in natural hybrid zones, as seen with monkey hybrids (*Callithrix*) in Brazil that are genetically differentiated in anthropogenic zones, with unpredictable outcomes for both the hybrids and the parental species [85].

    Another example of a successful hybrid is the feral hybrid pig resulting from the interbreeding of domestic and feral pigs. Feral pigs (*Sus scrofa*) are associated with many environmental impacts, such as species extinctions [86], alien introductions [87], and pathogen propagation [88]. Domestic pigs (*Sus scrofa domesticus*), on the other hand, have artificially selected traits, resulting in bigger individuals that generate more offspring [89]. Feral hybrid pigs can inherit both traits and produce larger litters and attain dense populations [90]. Those hybrid organisms transform ecosystems, and their interactions result in different evolutionary outcomes.

**Box 3:** *Genetically Modified Organisms (GMOs)*

Gene exchanges between species induced by humans may be one of the strongest life-changing mechanisms on Earth. The broad definition of GMOs includes transgenic organisms (where part of the genetic material of one species is transferred to another) [74]; cisgenic organisms (that have introduced genes originally from the same or a sexually compatible species) [91]; and epicrops (which have undergone epigenetic alterations involving agronomically important traits) [92].

Some transgenic organisms favor humans directly, such as providing vaccines and drugs [93] but others may bring unexpected or undesirable outcomes, such as the spread of new pests [94] and the increased mortality of non-target species [95]. Crop GMOs alter the ecosystem around them by changing natural processes and functions [96] – which leads to more questions than answers. There are uncertainties about the risks of transgene spread by hybridization and introgression events in areas bordering agrosystems [73]. In those cases, the transgenic hybrids could have higher fitness than parental organisms, depending on biotic and abiotic conditions [97], resulting in completely unpredictable and locally variable events that are essentially impossible to control in nature.

Cisgenic plants supposedly pose less risk to the environment because that type of gene transfer could happen in nature, although more experiments with cisgenic plants are needed to ensure their safety for commercial purposes [91].

Epigenetic alterations are naturally induced and regulated by the environment and have adaptive value, and epicrops carry epigenetic modifications induced by humans to favor agronomically important traits [98]. A drawback to epicrops resides in the distinct evolutionary dynamics of epigenetics, which occur more frequently and transitorily as compared to genetic alterations [99]. As a result, those plants are not yet commercially available.

The constant development and improvement of technologies to genetically modify organisms will generate even more uncertainties as GMOs spread. In the future, there could be even more widespread novel organisms linked to anthropogenic changes, transforming evolution along unexpected and unforeseen paths.

**Theoretical scenarios**

We are living the Great Acceleration [37], leading to changes in the landscape, organisms, humans, and habits. So, albeit unpredictable, the general outcomes of human-driven scenarios can be explored in relation to the expansion and retraction of anthropogenic novelty and native organisms. Although biotic and abiotic alterations of their habitat are usually pitfalls for native organisms, they could favor the establishment of alien species, hybrids, GMOs, and other novelty organisms in modified habitats.

We have outlined scenarios with the likely expansion or retraction of the global distribution of organisms grouped as native, alien, and anthropogenic-favored in the future based on two variables: a) environmental degradation and climate change (indirect human-driven changes), and b) human expansion and use of natural resources (direct human-driven changes) (Figure 1).

The more intense the indirect changes on the $x$ axis are, the more the alien species will spread – but not necessarily the anthropogenic-favored ones. The more intense the direct changes in the $y$ axis are, the more anthropogenic-favored organisms will spread (and possibly alien species too). Native organisms would decline in both situations of increases of human-driven changes (Figure 1).

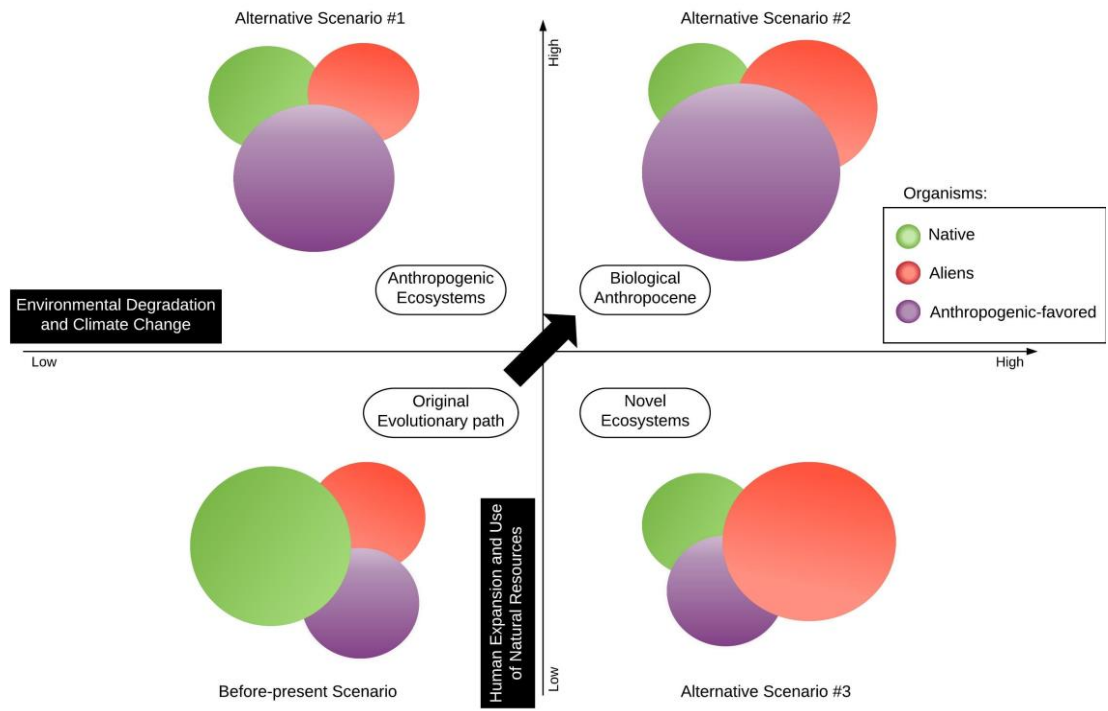

Figure 1: Theoretical global scenarios based on two variables for organisms grouped as native (green circles), alien species (red circles) and anthropogenic-favored (purple circles) showing the expansion and/or retraction of each group. Three alternative future scenarios were created based on two hypothetical variables: a) environmental degradation and climate change (x axis, indirect human-driven changes), and b) human expansion and use of natural resources (y axis, direct human-driven changes). The size of the circles corresponds to the abundances of the organisms on Earth within each scenario.

The three theoretical scenarios proposed here are simplifications of complex cause–consequence relationships and show possible outcomes of the *BioAnthro*. Alternative scenario #1 can be interpreted as representing anthropogenic ecosystems where there is high human interference in nature, but little environmental degradation or climate change (Figure 1). This scenario is a suitable prediction in case global climate change mitigations and habitat conservation are highly effective.

The *BioAnthro*, as we understand it, has both variables at high levels, with environmental degradation and climate change being irreparable, and the alterations and the use of natural resources through human hyper-dominance are immeasurable yet astonishingly high in alternative scenario #2 (Figure 1). In the alternative scenario #3, the environmental degradation is high but human interference is low. This could be similar to the novel ecosystem concept, where the spread of alien species is expected with moderate changes in biodiversity (Figure 1).

So, based on the current level of environmental disturbance and transformation, we believe we are moving quickly and directly to scenario #2, the *Biological Anthropocene*, where the original evolutionary paths are completely shifted, with the spread of alien species and anthropogenic-favored organisms and the decline of native organisms. Briefly, in future ecosystems, there will be little space for native organisms with no economic use, while alien species and anthropogenic-favored organisms will spread in disorderly fashions, creating novel interactions and novelty organisms along unexpected evolutionary paths.

**Conclusions**

Humans are impacting life and the environment on Earth in unprecedented ways. The advance of technology and the maintenance of diverse human habits have created new evolutionary pressures on all living organisms. Some of the disturbances created by human hyper-dominance over all of Earth's systems can lead to losses of biodiversity. However, other anthropogenic changes include novelty organisms that are constantly being introduced and incorporated into all ecosystems, permanently transforming evolution.

These evolutionary novelty organisms are alien species, hybrids, GMOs and other organisms either created or induced by human habits, cultures and technology. The functional boundaries limiting anthropogenic ecosystems from natural ones are very permeable and hard to identify. Thus, those organisms will interact with all others, intentionally and unintentionally, thereby changing evolutionary pathways. The interactions among all organisms – the novel, the introduced, and the native – are constantly changing and adapting through additive and fractal relationships, with unexpected and unforeseen outcomes, called here the *Biological Anthropocene*.

The *BioAnthro* incorporates uncertainties about the future of Earth's biota, as novel interactions are unavoidable. The changes introduced by technology and culture are now exponential due to human hyper-dominance. Therefore, to understand the outcomes of those interventions, it will be necessary to consider and rethink humans as crucial modifying agents of evolution and novelty organisms (such as alien species, hybrids, and GMOs) as transforming agents, all shifting evolutionary pathways.

**Acknowledgements**

Financial support was provided by Ministério do Meio Ambiente/PROBIO II. We thank XX reviewers for comments on the manuscript, Mr. Roy Funch for linguistic advice.